\documentclass[english]{paper}
\usepackage[T1]{fontenc}
\usepackage[latin9]{inputenc}
\usepackage{float}
\usepackage{units}
\usepackage{amsmath}
\usepackage{amsthm}
\usepackage{amssymb}
\usepackage{graphicx}
\usepackage{cite}

\makeatletter
\numberwithin{equation}{section}
\numberwithin{figure}{section}

\usepackage{tikz}
\usepackage{tikz-3dplot}
\usetikzlibrary{arrows,external}
\usepackage{pgfplots}
\pgfplotsset{compat=1.3}
\usepackage[detect-family]{siunitx}
\usepackage[eulergreek]{sansmath}
\usepackage{relsize}
\usepgfplotslibrary{polar}

\makeatother

\usepackage{babel}

\begin{document}

\title{The field of view of a scintillator pair for cosmic rays}

\author{N.G. Schultheiss\thanks{Nikhef} \thanks{Zaanlands Lyceum}}

\maketitle

\begin{abstract}
Particles in an extended air shower (EAS), initiated by a cosmic ray
primary, lead to two nearly simultaneous detections in a scintillator
pair. The angle of the EAS and the axis through both scintillators
can be reconstructed using the time difference of the detections and
the distance between the scintillators. The acceptances of a scintillator
along the axis through the scintillators and perpendicularly on this
axis follow the same distribution in theory. Using a data set with
two perpendicular detector pairs this theory is verified. The distribution
of possible origins of cosmic ray primaries, and the resulting EAS,
can thus be described using the perpendicular distribution for a given
time difference.\end{abstract}

\section{Introduction}

Cosmic ray primaries initiate extended air showers (EAS) in the atmosphere.
The number of secondary particles in an EAS depends both on the energy
and the type of the cosmic ray primary particle (foton, electron,
proton, nucleus). Along the path in the atmosphere the energy in the
EAS is distributed over a growing number of particles. With sufficient
energy of the primary particle ($E_{p}>\unit[10^{14}]{eV}$) a cascade
of secondary particles can reach scintillators on the Earth's surface
and are detected. Comparing the zenith angles of cosmic ray primaries
with equal energy, an increasing zenith angle leads to an increasing
path length for the resulting EAS in the atmosphere. As a result the
number of detectable particles that reach the surface will decrease.

In HiSPARC the direction of a cosmic ray primary penetrating the Earth's
atmosphere is reconstructed using measurements of EAS particles in
scintillators. This is simulated for a single detector using $10^{6}$
primary particles having random distributions for the zenith angle
$\theta$ and the azimuth $\varphi$. The distribution of $dN\left(\varphi\right)/d\varphi$
is flat, because of the symmetry properties around the zenith. The
distribution $dN\left(\theta\right)/d\theta$ is described as \cite[p. 78]{G-PCRP}:

\begin{equation}
\frac{dN\left(\theta\right)}{d\theta}\varpropto2\pi\sin\left(\theta\right)\cos^{a}\left(\theta\right)\label{eq:rossi}
\end{equation}

This distribution has a geometrical and a physical component. The
geometrical component can be written as $2\pi\sin\left(\theta\right)\cos\left(\theta\right)$.
The factor $2\pi\sin\left(\theta\right)$ is proportional to the celestial
area and $\cos\left(\theta\right)$ is proportional to the effective
area of the detector perpendicular to the displacement of the EAS.
The physical component due to the extinction in the atmosphere is
described as being proportional to $\cos^{\left(a-1\right)}\left(\theta\right)$. 

\begin{figure}
\noindent \begin{centering}
\includegraphics[width=8cm]{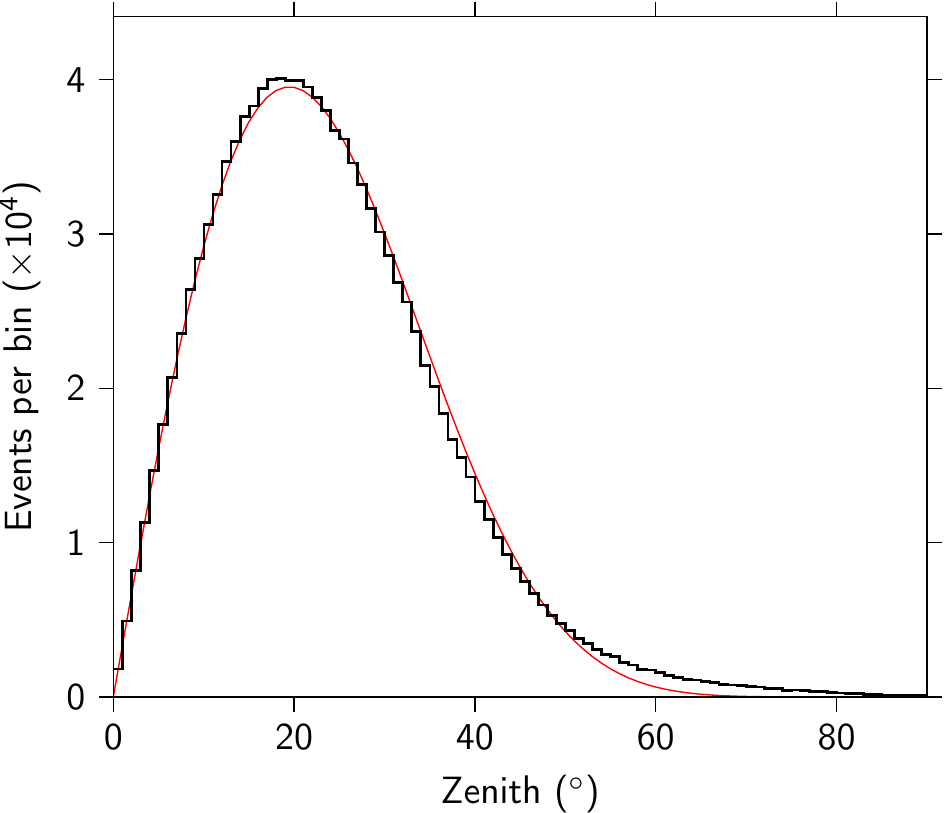}
\par\end{centering}

\caption{\textit{\label{fig:The-measured-number}The measured number of particles
as function of the zenith angle $\theta$ for Science Park Amsterdam
shown as a step histogram. The fitted distribution with $a=7$ is
plotted as a red curve. Diagram from} \cite{AdL}}
\end{figure}

\begin{figure}[h]
\begin{centering}
\includegraphics[width=10cm]{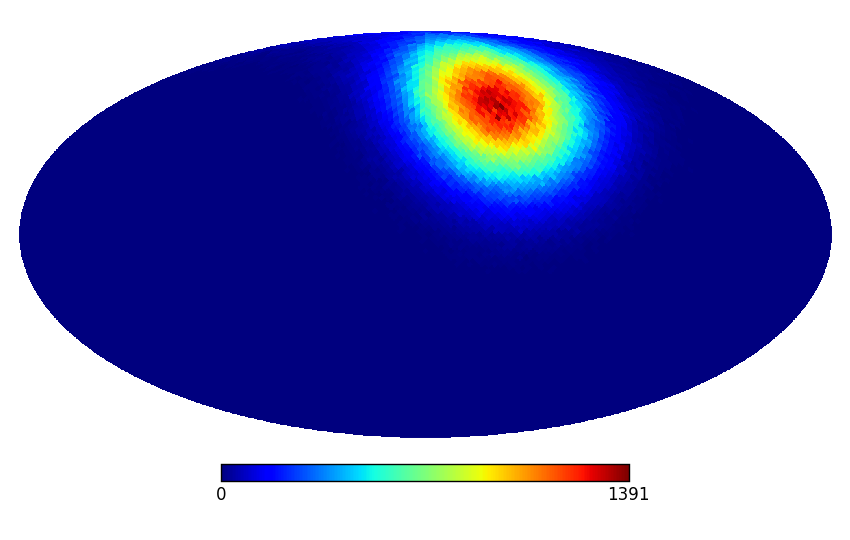}
\par\end{centering}

\caption{\label{fig:singleScintillatorSimulation}\textit{A simultion of $10^{6}$
EAS shown in a Mollweide view (equal area bins) neglecting the Earth's
rotation. The number of EAS per unit of area is shown in an equatorial
projection. The north or Polaris is in the highest point, south is
the lowest point. The zenith of the scintillator is at the center
of the colored area.}}
\end{figure}

The location in space of the detector is time dependent as a result
of the rotation of the Earth. This rotation of the Earth is neglected
in the simulation shown in fig. \ref{fig:singleScintillatorSimulation}\footnote{The 
corresponding atmospheric layer converts a cosmic primary in an
EAS and is regarded as a part of the detection setup.}. The majority of the detected 
particles is caused by the abundant cosmic ray primaries in the range of 
$\unit[10^{14}]{eV}$. The distribution of directions of arrival for cosmic rays with this 
energy is very uniform \cite[p. 124-125]{G-PCRP}. In this case
the acceptance is considered independent on the azimuth.

\begin{figure}[H]
\noindent \begin{centering}
\includegraphics[width=10cm]{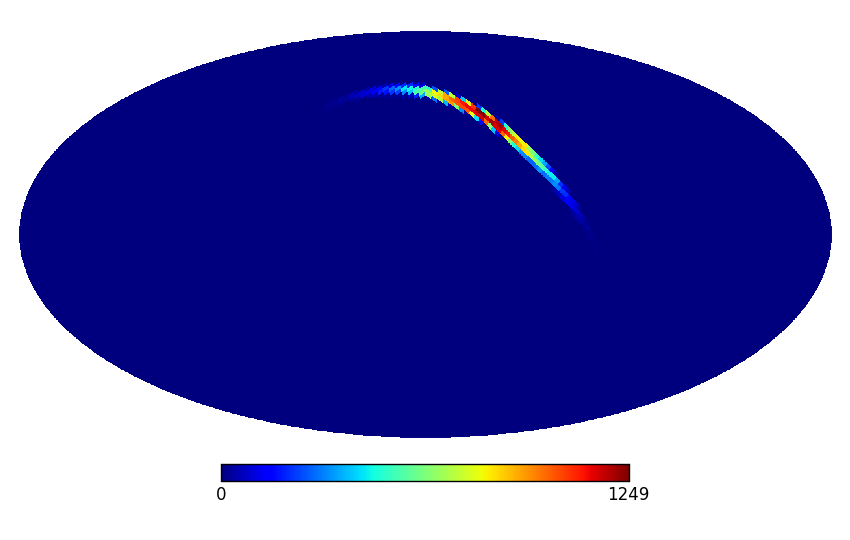}
\par\end{centering}

\caption{\label{fig:pairSimulation}\textit{The dataset of fig. \ref{fig:singleScintillatorSimulation}
is used for the simulation of a station with two detectors. The line
connecting the centers of the two scintillators makes an angle of
$45^{\mathrm{o}}$ with respect to the local meridian. The time difference
between detections is }5ns\textit{ with a bin width of $2.5$}ns\textit{.
The distance between the scintillators is }10m\textit{. }}
\end{figure}

\newpage{}

Discrimination between background radiation and EAS particles is not
possible using a single scintillator. A cosmic ray event is generated when both
scintillators detect particles. In this case the EAS spreads particles over an
area covering both scintillators. Background radiation only generates
a detection in a single scintillator.

An added advantage of a setup using a scintillator pair is the possibility
to partially reconstruct the direction of the shower axis (fig. \ref{fig:pairSimulation}).
The particles in an EAS travel at speeds approximating the speed of
light $c$. The time difference $t_{1,2}$ of the particle detections
and the distance $d$ between the scintillators define the angle $\theta_{dt}$
of that EAS:

\begin{equation}
\sin\left(\theta_{dt}\right)=\frac{c\,t_{1,2}}{d}\label{eq:zenith_reconstruction}
\end{equation}

\section{Theory}

\subsection{Distributions}

\begin{figure}
\noindent \begin{centering}
\includegraphics[width=7cm]{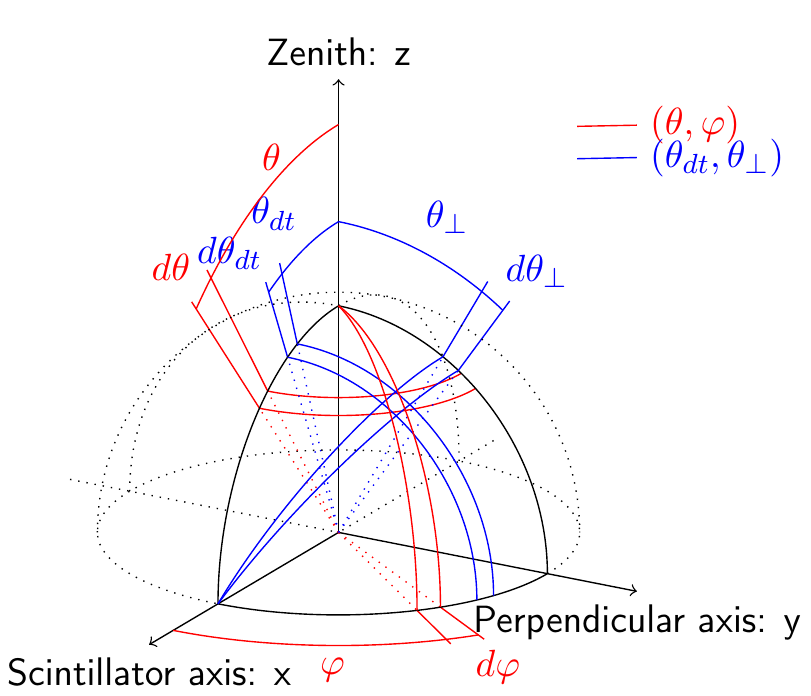}
\par\end{centering}

\caption{\label{fig:celestialArea}\textit{Two intersecting area's with the
same location are described using two polar coordinate systems.} \textit{The
$\left(\theta,\varphi\right)$ coordinate system is symmetric around
the zenith (shown in red). The $\left(\left(\pi/2-\theta_{dt}\right),\theta_{\perp}\right)$
coordinate system is symmetric around the axis through the scintillators
(shown in blue). }}
\end{figure}

In fig. \ref{fig:celestialArea} the direction of an EAS is shown
as the intersection of a slice and a wedge on a celestial sphere.
This intersection usually is expressed in the $\left(\theta,\varphi\right)$
or zenith, azimuth polar coordinate system. The sky is symmetric around
the axis defined by the zenith. This symmetry leads to equal distributions
in slices trough the zenith.

\begin{figure}
\noindent \begin{centering}
\includegraphics[width=7cm]{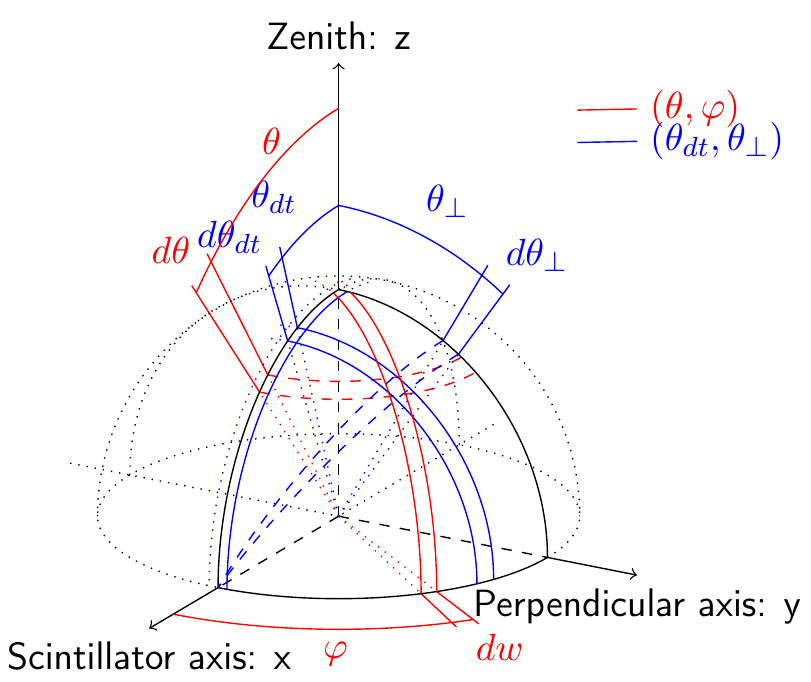}
\par\end{centering}

\caption{\label{fig:celestialArea-1}\textit{An area described using the intersections
of slices of equal width.} \textit{The defined area can be reached
in two separate ways. Along the red slice the flux can be calculated
immediately using $\theta$. The same location is reached in two steps
along the blue slices. Along the blue slice trough the scintillator
axis the perpendicular slice is reached. The flux is calculated at
$\theta_{dt}$, defining the number of particles in the perpendicular
slice. The flux of this second set of particles is distributed along
the perpendicular slice as a function of $\theta_{\perp}$ in a similar
way. }}
\end{figure}

The axis through both scintillators defines a symmetry in a detection
station. Here a $\left(\left(\pi/2-\theta_{dt}\right),\theta_{\perp}\right)$
polar coordinate system is used\footnote{It must be noted that the atmospheric depth is dependent on $\theta_{dt}$
and $\theta_{\perp}$, this effects the number of particles in the
EAS.}. The angle $\theta_{dt}$ is reconstructed using the time difference
in the detections (eq. \ref{eq:zenith_reconstruction}). Because of
the symmetry, $\theta_{\perp}$ is not directly determinable.

Using eq. \ref{eq:rossi} the number of EAS coming from a circular
band (slice) with length $2\pi\sin\left(\theta\right)$ and width
$d\theta$ can be calculated. In figure \ref{fig:celestialArea} a
segment of the wedge of the celestial half sphere defined by $\varphi$
and $\varphi+d\varphi$ intersects with a slice defined by $\theta$
and $d\theta$ is shown in red. The EAS-flux in the intersecting area
is formulated as:

\begin{equation}
\frac{d^{2}N\left(\theta,\varphi\right)}{d\theta d\varphi}\varpropto\sin\left(\theta\right)\cos^{a}\left(\theta\right)
\end{equation}

The width of the wedge is expressed as $\sin\left(\theta\right)d\varphi$.
The equal width $dw$ of a slice, as shown in figure \ref{fig:celestialArea-1},
is not dependent on $\theta$. This leads to a flux of:

\begin{equation}
\frac{d^{2}N\left(\theta,\varphi\right)}{d\theta dw}\varpropto\cos^{a}\left(\theta\right)\label{eq:distribution_zenith}
\end{equation}

Eq. \ref{eq:distribution_zenith} is a valid expression for all slices
with constant width and through the zenith due to the symmetry around
the zenith axis.

The location of an intersection on a unit sphere can be written in
the $\left(\theta,\varphi\right)$-frame as:

\begin{equation}
\left(\begin{array}{c}
x\\
y\\
z
\end{array}\right)=\left(\begin{array}{c}
\sin\left(\theta\right)\cos\left(\varphi\right)\\
\sin\left(\theta\right)\sin\left(\varphi\right)\\
\cos\left(\theta\right)
\end{array}\right)
\end{equation}

And in the $\left(\theta_{dt},\theta_{\perp}\right)$-frame as:

\begin{equation}
\left(\begin{array}{c}
x\\
y\\
z
\end{array}\right)=\left(\begin{array}{c}
\cos\left(\pi/2-\theta_{dt}\right)\\
\sin\left(\pi/2-\theta_{dt}\right)\sin\left(\theta_{\perp}\right)\\
\sin\left(\pi/2-\theta_{dt}\right)\cos\left(\theta_{\perp}\right)
\end{array}\right)=\left(\begin{array}{c}
\sin\left(\theta_{dt}\right)\\
\cos\left(\theta_{dt}\right)\sin\left(\theta_{\perp}\right)\\
\cos\left(\theta_{dt}\right)\cos\left(\theta_{\perp}\right)
\end{array}\right)
\end{equation}

The $z$-coordinate expresses $\theta$ as a function of $\theta_{dt}$
and $\theta_{\perp}$:

\begin{equation}
z=\cos\left(\theta\right)=\cos\left(\theta_{dt}\right)\cos\left(\theta_{\perp}\right)\label{eq:angle_result}
\end{equation}

Substition of eq. \ref{eq:angle_result} in eq. \ref{eq:distribution_zenith}
leads to:

\begin{equation}
\frac{d^{2}N\left(\theta,\varphi\right)}{d\theta dw}\varpropto\cos^{a}\left(\theta_{dt}\right)\cos^{a}\left(\theta_{\perp}\right)
\end{equation}

Leading to:

\begin{equation}
\frac{d^{2}N\left(\theta,\varphi\right)}{d\theta dw}\varpropto\frac{d^{2}N\left(\theta_{dt},w\right)}{d\theta_{dt}dw}\times\frac{d^{2}N\left(\theta_{\perp},w\right)}{d\theta_{\perp}dw}\label{eq:theory}
\end{equation}

The mathematical analysis proofs that all distributions in the slices
in fig. \ref{fig:celestialArea-1} follow the same function.

\section{\label{sec:Verification}Verification}

\begin{figure}[H]
\vspace{0.5cm}

\noindent \begin{centering}
\includegraphics[width=7cm]{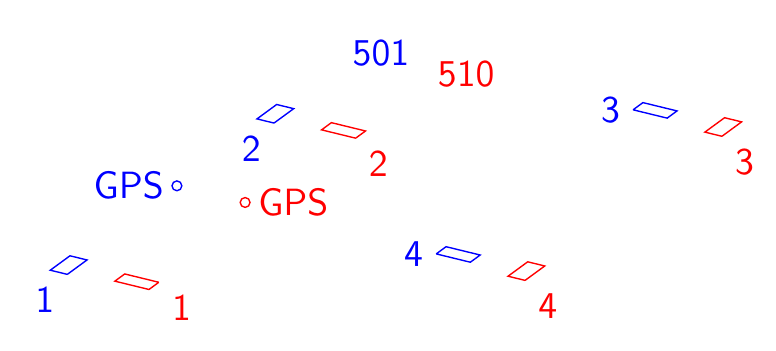}
\par\end{centering}

\vspace{0.5cm}

\caption{\label{fig:stationMap}\textit{The map of stations $501$ and $510$.}
\textit{Station $\mathrm{501}$ is shown in blue, $\mathrm{510}$
is shown in red. A station consists of four scintillators $(1$, $2$,
$3$ and $4)$ with areas of $1000\times500\mathrm{mm^{2}}$ and a
$\mathrm{GPS}$ antenna used for timing. The axes defined by detector
pairs $2$ and $4$ are perpendicular to the axes defined by pairs
$1$ and $3$. Distances $d_{1,2}$, $d_{2,3}$, $d_{3,4}$, $d_{4,1}$
and $d_{2,4}$ are all $10\mathrm{m}$. Distance $d_{1,3}$ is $17.3\mathrm{m}$.
Distances between similar points in $501$ and $510$ are $2\mathrm{m}$.}}
\end{figure}

Stations 501 and 510 both contain four scintillators located on perpendicular
axes as shown in fig. \ref{fig:stationMap}. Each scintillator is
equipped with a Photo Multiplier Tube (PMT). Two PMT's are connected
via cables to a single HiSPARC electronics unit. A master and a slave
unit register all events from two scintillator pairs in one station
with a sample frequency of 400MHz with a constant delay for each scintillator.
The generated timing errors, due to these constant delays, must be
compensated in the algorith.

The resulting data is sent via a measurement computer to the HiSPARC
data repository. Both stations have independent clocks in the master
and slave units. These are synchronized using a GPS module in each
station. Timing errors generated by this setup can be compensated
because the cosmic ray flux from the zenith is at a maximum, leading
in this case to equal arrival times for particles in both scintillators. 

\begin{figure}[H]
\noindent \begin{centering}
\includegraphics[width=8.0cm]{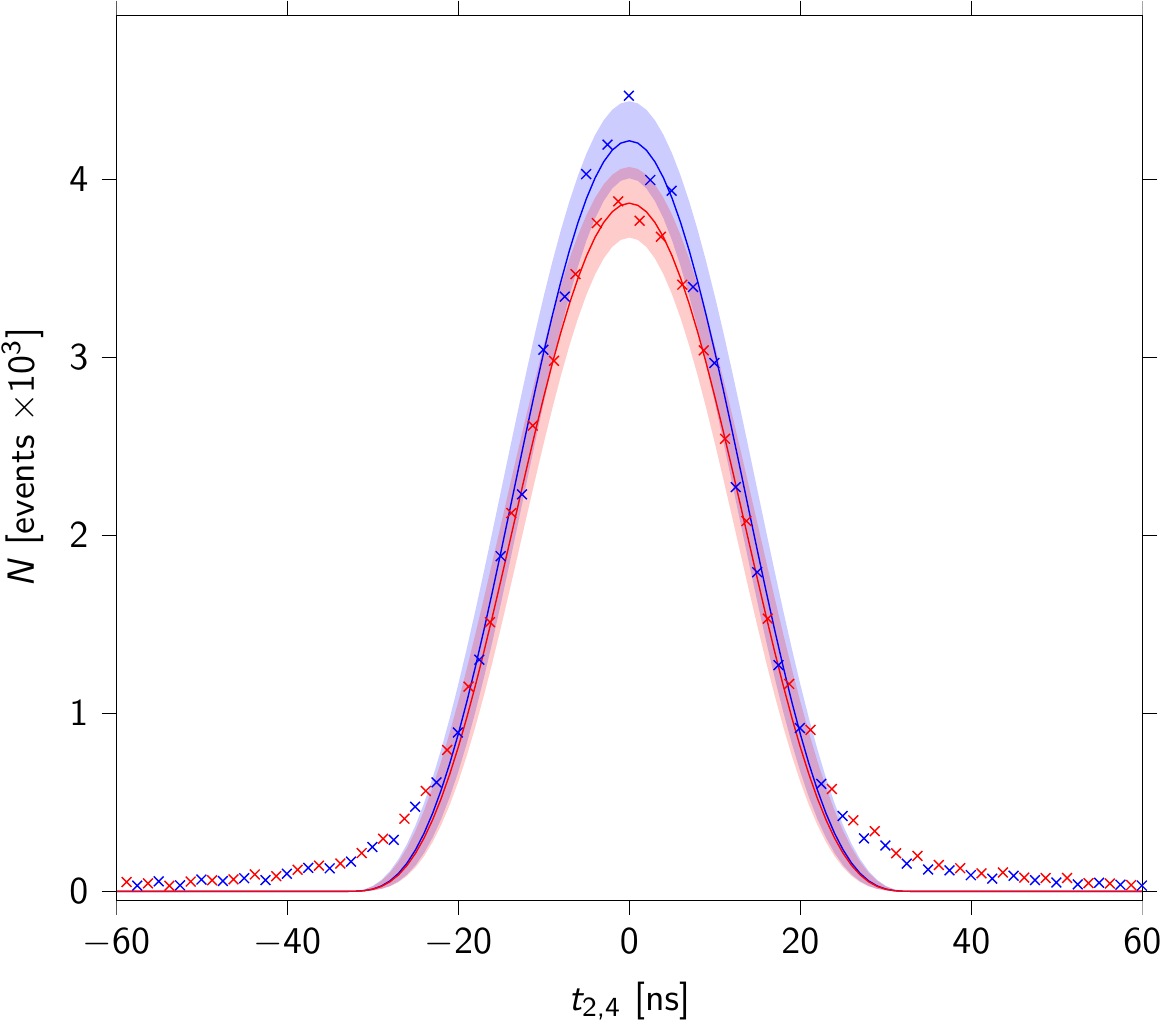}
\par\end{centering}

\caption{\label{fig:pairDistribution-501-510-2/4}\textit{Distributions along
the axis from scintillator 2 to 4 are shown.} \textit{Data in the
period from 5 October 2014 to 25 March 2015 using stations $501$
(blue) and $510$ (red) is binned in $\unit[2.5]{ns}$ wide bins and
shown as crosses. The distributions for scintillator pairs 2 and 4
with distances of $10\mathrm{m}$ are fitted. Shown distributions
are calculated using $a=7\pm1$.}}

\noindent \begin{centering}
\includegraphics[width=8.0cm]{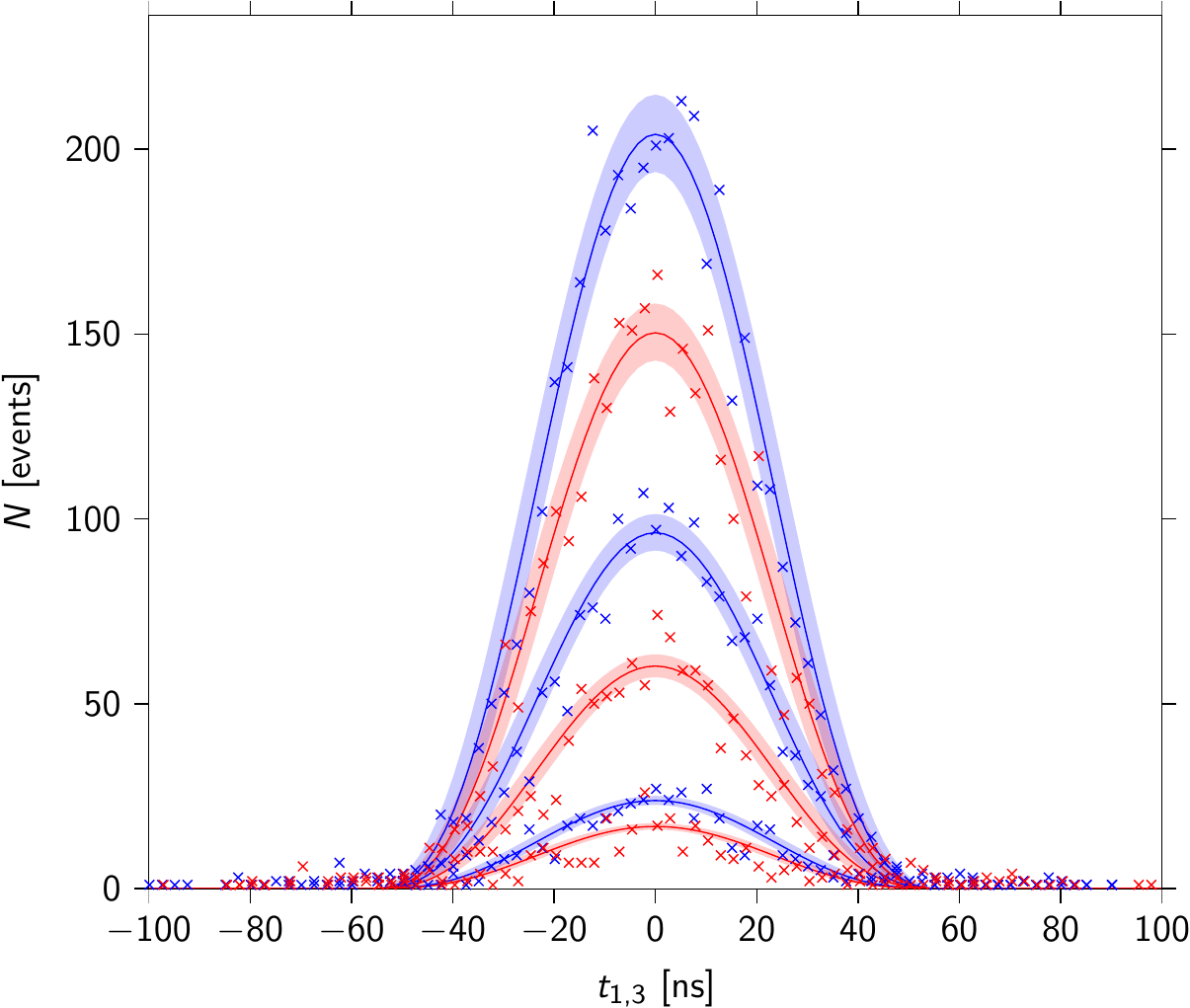}
\par\end{centering}

\caption{\label{fig:pairDistribution-501-510-1/3}\textit{Distributions for
the perpendicular pairs are shown. Measured particle showers are binned
for time difference $\triangle t_{2,4}$, along the parallel axis
from scintillator 2 to scintillator 4. The distribution along the
perpendicular axis, from scintillator 1 to scintillator 3, is shown
for three time differences ($\triangle t_{2,4}$) $10\mathrm{[ns]}$
apart ($501$ is shown in blue and $510$ in red). Shown distributions
are calculated using $a=7\pm1$.}}
\end{figure}

Scintillator pairs 2 and 4 (for both station 501 and 501) are regarded
as single scintillator pairs. Scintillator pairs 1 and 3 (of both
stations) are used to verify the theory of equal distributions. A
data set over the periode from 5 October 2014 to 25 March 2015 is
used. This set contains events where all scintillators have detected
particles. Fig. \ref{fig:pairDistribution-501-510-2/4} shows the
measured distributions for the stations 501 and 510 for scintillators
2 and 4. Both distributions are compensated for systematic errors. 

Using eq. \ref{eq:distribution_zenith} a curve fit is generated,
leading to the exponent $a=7\pm1$, the error along the $N$-axis
is set on 5\%.

To verify the theory, the data set is divided into subsets. Each subset
is selected on a binned time difference for the scintillators 2 and
4. If the time difference is equal to the timing error, a slice through
the zenith is selected. In fig. \ref{fig:pairDistribution-501-510-1/3}
the distributions for scintillator pairs 1 and 3 are shown for selected
subsets with time differences 10ns apart. The measurements of each
subset fit the plotted distribution with exponent $a=7$ with an error
along the $N$-axis of 5\%.

Both plots in fig. \ref{fig:pairDistribution-501-510-2/4} and fig.
\ref{fig:pairDistribution-501-510-1/3} show a discrepancy for larger
angles. This partly can be explained by the size of the scintillators.
These have an area of $1000\times500\mathrm{mm^{2}}$. The distance
between arriving particles is between $9.25\mathrm{m}$ and $10.75\mathrm{m}$
or 16.55m and 18.05m because paired scintillators are placed perpendicular
towards each other. This results in an extra error of $7.5\%$ for
a scintillator distance of $10\mathrm{m}$ and $4.5\%$ (fig. \ref{fig:pairDistribution-501-510-2/4})
for a scintillator distance of $17.3\mathrm{m}$ (fig. \ref{fig:pairDistribution-501-510-1/3}).
Fig. \ref{fig:The-measured-number} shows a simular discrepancy between
measurements and the fitted curve.

\section{Reconstruction}

A reconstrucion of the field of view of a scintillator pair starts
with the measured arrival times $t_{m,1}$ and $t_{m,2}$ a time difference
$t_{m,1\rightarrow2}=t_{m,2}-t_{m,1}$ is calculated. The setup as
explained in sec. \ref{sec:Verification} generates constant delays
for each scintillator leading to a timing error $t_{e,1\rightarrow2}$.
The corrected time difference $t_{1,2}$ is calculated using the timing
error $t_{e,1\rightarrow2}$: $t_{1,2}=t_{m,1\rightarrow2}-t_{e,1\rightarrow2}$.
This corrected time difference of the arriving particles is used to
reconstruct $\theta_{dt}$. 

\begin{figure}[h]
\noindent \begin{centering}
\includegraphics[width=8.0cm]{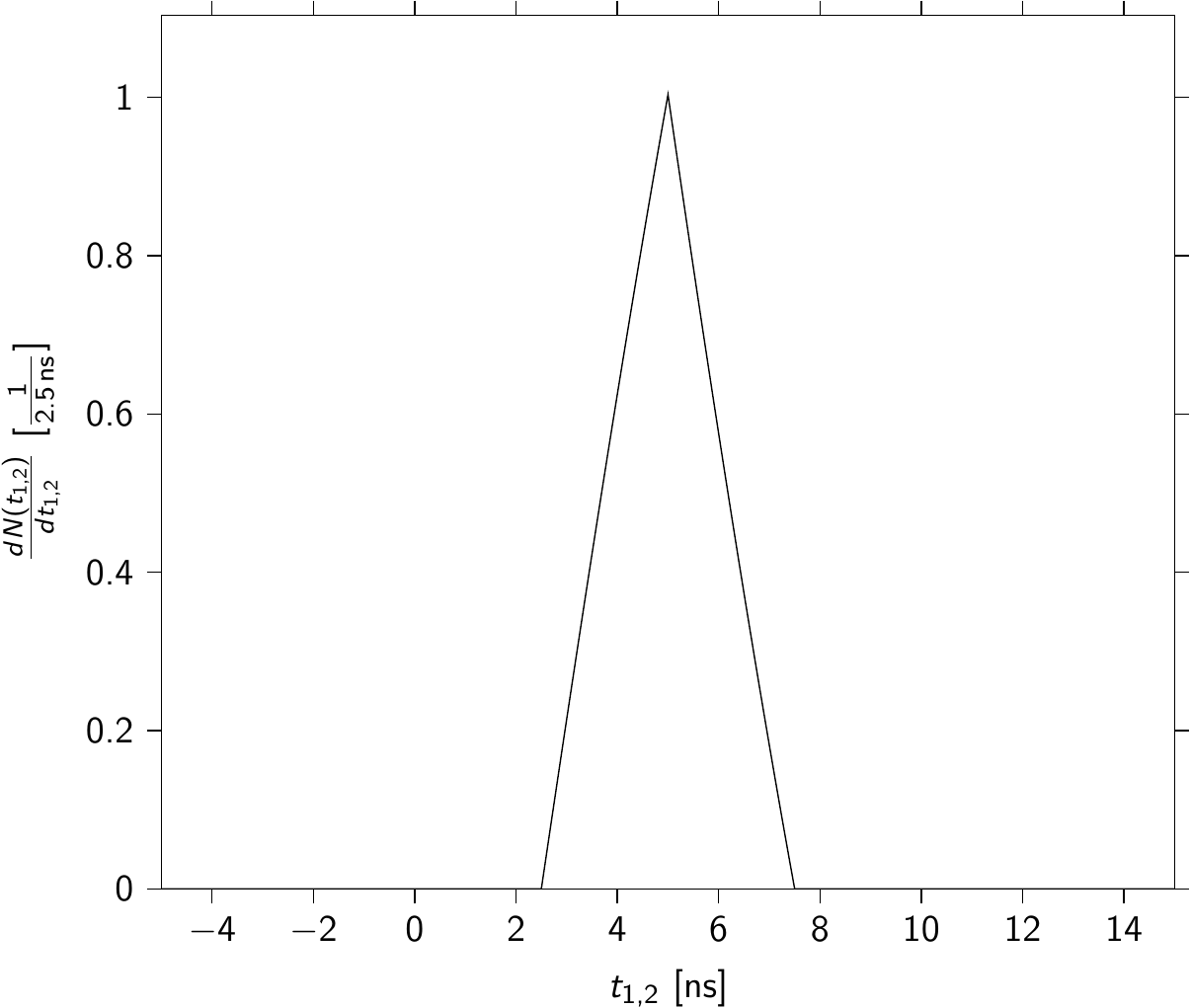}
\par\end{centering}

\noindent \centering{}\caption{\label{fig:time-distribution}\textit{The probability distribution
is at a maximum when a time difference is an exact measured binning
time : ..., $-2.5\mathrm{ns}$, $0.0\mathrm{ns}$, $2.5\mathrm{ns}$,
... . There is a drop off when the difference moves to the bin edges.
The sides of the triangle are slightly curved due to the particle
distribution through the zenith.}}
\end{figure}

The angle $\theta_{dt}$ is calculated using eq. \ref{eq:zenith_reconstruction}:

\begin{equation}
\theta_{dt}=\arcsin\left(\frac{c\,t_{1,2}}{d}\right)
\end{equation}

The sample frequency of \unit[400]{MHz} leads to a binwidth of \unit[2.5]{ns}.
The exact time difference is distributed over an interval of two measured
bins. If the first time is close to the begin of the first bin and
the second time is close to the end of the second bin there is a small
chance of an extra bin time difference (and vice versa). This results
in a triangular probability distribution $dn\left(t\right)/dt$. 

Eq. \ref{eq:distribution_zenith} defines the number of particles
in a binned slice with a time difference.

\begin{equation}
\frac{d^{2}N\left(t_{1,2},w\right)}{dt_{1,2}dw}\varpropto\cos^{a}\left(\arcsin\left(\frac{c\,t_{1,2}}{d}\right)\right)
\end{equation}

\begin{figure}[h]
\noindent \begin{centering}
\includegraphics[width=8.0cm]{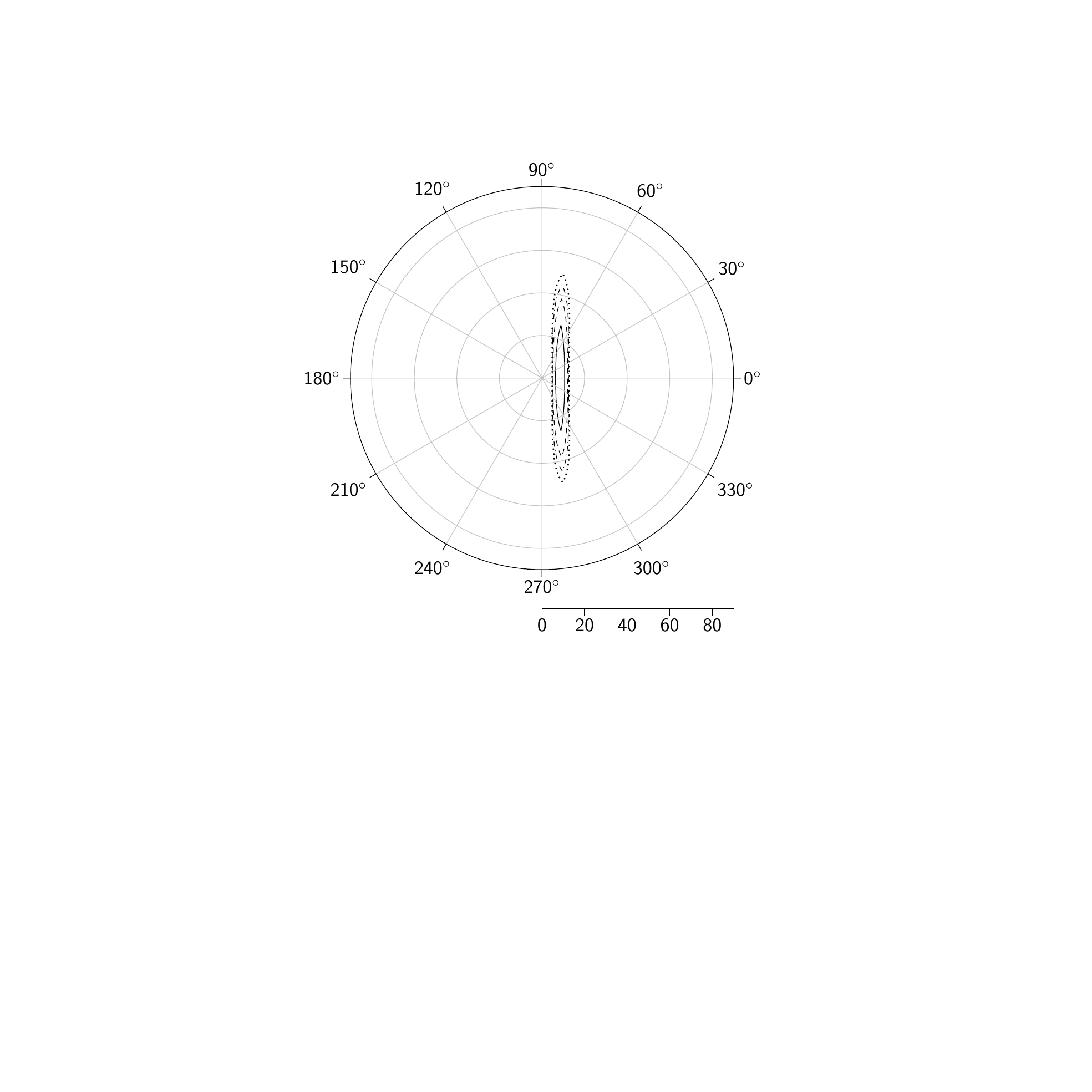}
\par\end{centering}

\caption{\label{fig:Field-of-view}Field of view for $t_{1,2}=5\mathrm{ns}$
\textit{The field of view is shown in a $\left(\theta,\varphi\right)$-plot
for a detection with $t_{1,2}=5\mathrm{ns}$ and a scintillator distance
of $10\mathrm{m}$, $\theta$ is plotted along the radius. The curves
correspond with probabilities of $5\%$ (dotted), $10\%$ (dash dotted),
$20\%$ (dashed) and $50\%$ (solid). The scintillators are locared
on the axis $0^{\mathrm{o}}$ / $180^{\mathrm{o}}$, as a result the
angle $\theta_{dt}$ is along the $0^{\mathrm{o}}$ / $180^{\mathrm{o}}$
axis and the angle $\theta_{\perp}$ is along the $90^{\mathrm{o}}$
/ $270^{\mathrm{o}}$ axis.}}
\end{figure}

\newpage{}

Using the triangular distribution $dn\left(t\right)/dt$ the number
of particles is calculated for a given time. 

\begin{equation}
\frac{d^{2}N\left(t,w\right)}{dtdw}=\frac{dN\left(t\right)}{dt}\times\frac{d^{2}N\left(t_{1,2},w\right)}{dt_{1,2}dw}
\end{equation}

This resulting distribution of time differences\footnote{Normalised for a single particle.}
is shown in fig. \ref{fig:time-distribution} for a corrected time
difference of $t_{1,2}=\unit[5.0]{ns}$. The distribution of time
differences follows a triangle, slightly deformed due to the curvature
of the particle distribution as a function of the time difference. 

The angle $\theta_{\perp}$ now can be calculated for a given chance.
Fig. \ref{fig:Field-of-view} shows these curves for several chances
defining the field of view for a time difference in detections.

\section{Conclusion}

A single scintillator has a wide field of view, a scintillator pair
has a an increased accuracy along the axis through both scintillators.
Mathematical analysis shows that all distributions of particles in
planes perpendicular to the Earth's surface are isomorphous. The time
difference defines a cone of arrival for particles in an EAS. Distributions 
in this cone and perpendicular on this cone are simular. 
This is verified using four scintillators placed on perpendicular axes in the 
HiSPARC stations 501 and 510. The size of the scintillators and the 
thickness of the EAS front are not taken into account. Discrepancies between
theory and measurements are however small.

Using this theory the field of view of a scintillator pair can be
reconstructed using the time difference. Comparing the theoretical
field of view with the simulation of a detector pair both show an
elongated band across the sky.

The author wishes to thank J.J.M. Steijger for many critical discussions.

\bibliographystyle{plain}
\bibliography{references}

\end{document}